# Flexible-AR display for near-eye operations

Alan Lee
Department of Electrical and Electronic Engineering
The University of Melbourne
Melbourne Australia
alanl@unimelb.edu.au

Dechuan Sun
Department of Electrical and Electronic Engineering
The University of Melbourne
Melbourne Australia
dechuan.sun@unimelb.edu.au

Gregory Tanyi
Department of Electrical and Electronic Engineering
The University of Melbourne
Melbourne Australia
gregory.tanyi@unimelb.edu.au

Younger Liang
*Division of Augmented Reality*
*KDH Design*
Taipei, Taiwan
younger@kdh-design.com

Christina Lim
Department of Electrical and Electronic Engineering
The University of Melbourne
Melbourne Australia
chrislim@unimelb.edu.au

Ranjith R. Unnithan
Department of Electrical and Electronic Engineering
The University of Melbourne
Melbourne Australia
r.ranjith@unimelb.edu.au

*Abstract*—We propose a new technique to fabricate flexible-near-field Argument-Reality (AR) display using modular-molds. A near-eye flexible-AR-display is fabricated based on parameters extracted from simulations. Our AR-display successfully reconstructed images and videos from a light-engine. It opens a new approach to fabricate flexible-near-field AR display with good physical stress and collision-resilience.

*Keywords—AR, AR display, near field, modular fabrication.*

## I. Introduction

Augmented Reality (AR) is a technological innovation that enables real-time user interaction with virtual data, demonstrating versatility across numerous domains, including education, navigation, entertainment, healthcare, and engineering [1-2]. This cutting-edge technology superimposes computer generated content onto the real world, resulting in a hybrid environment that combines both virtual and physical elements. Hence, it has the potential to become the next generation interactive and computing platform [3-5]. In the design and fabrication processes of conventional AR near-eye displays, the optical combiner emerges as the most challenging component due to various factors, such as issues related to image alignment, miniaturization, field of view expansion, color fidelity, optical aberrations, and cost considerations [4]. Based on their working principles, the commercially available combiners can be categorized into geometric and diffractive waveguide based combiners [4, 6].

The manufacturing process for diffractive waveguides shows substantial flexibility, requiring only the deposition of a thin film of materials onto the glass waveguide substrate. However, the diffractive waveguides demonstrate several limitations, among which chromatic aberration, often referred to as the rainbow effect, is particularly noteworthy [7], and the diffractive waveguides will lose more light efficiency than the geometrically coupled waveguides based on reflective architecture. Additionally, diffractive waveguides do not provide an expansive field of view comparable to that of geometric waveguides. These drawbacks primarily arise from the inherent sensitivity and selectivity of diffraction gratings [8,9] with respect to incident light input angles and wavelengths. They are challenging to overcome [10]. Although geometric waveguides do not exhibit these particular issues, the standard manufacturing process for geometric waveguides involves precise procedures, including molding, dicing, bonding, and polishing, all of which result in a low overall yield.

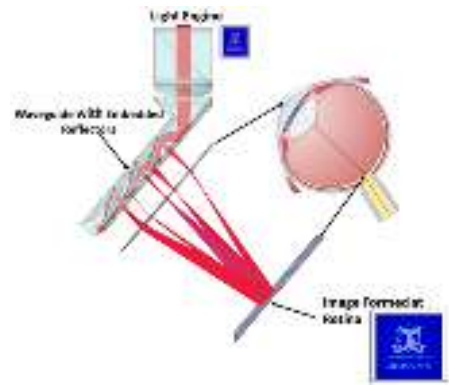

Fig. 1: 3-D schematic illustrating the light path, image combination, and pupil expansion as light propagates from coupling optics through the waveguide to the end user's eye.

In this paper, a novel molding method based on polyjet 3D printing techniques is proposed for the first time to fabricate geometric flexible AR waveguide combiners. Instead of employing a glass substrate, we utilized a flexible material polydimethylsiloxane (PDMS). The flexibility can enhance the waveguide's resilience to physical stress and collisions, a valuable feature in scenarios such as cycling or skiing where safety is of utmost importance, and the breakage of glass may pose a risk to the eyes. To the best of our knowledge, no flexible geometrically coupled AR waveguide displays have been shown previously. Furthermore, our proposed manufacturing method is cost-effective.

## II. Design of the PDMS AR display

Figure 1 depicts our proposed flexible near-field AR display system. PDMS is a flexible material that commonly employed for microfluidic fabrication. PDMS possesses a refractive index close to that of glass, approximately 1.43, making it a promising candidate for our display. Additionally, the hardness of PDMS can be fine-tuned by adjusting the curing hardener ratio [11]. The suggested dimensions for the active region of the AR display are 3mm x 45mm x 35mm, closely resembling those of a standard prescription eyeglass frame.

To practically realize the reflectors, we propose to embed dielectric-coated glass-slips in the PDMS during the fabrication as shown in Fig. 2 (a) and (b). To achieve the desired characteristics for our near-field AR display, including a large field of view (FOV), adequate transparency, and a wide eyebox, we employ three reflectors. In our proposal, PDMS serves as the optical light-guide material. Finite element method implemented in COMSOL Multiphysics [12] was used to conduct simulations of our device architecture. The Ray Optics module was used to determine the optimal angles for reflector tilt, thickness of the waveguide and input facet orientation for a perpendicular light engine configuration. These optimized angles are found to be 25 degrees and 50 degrees, respectively and the thickness to be 3 mm.

For our prototype, we employ dielectric-coated glass slips with a reflectivity of 10%. This choice is made to ensure that the AR waveguide meets the transparency requirement.

### III. Fabrication of near-field AR display using PDMS

Another advantage of using PDMS is the relatively simple

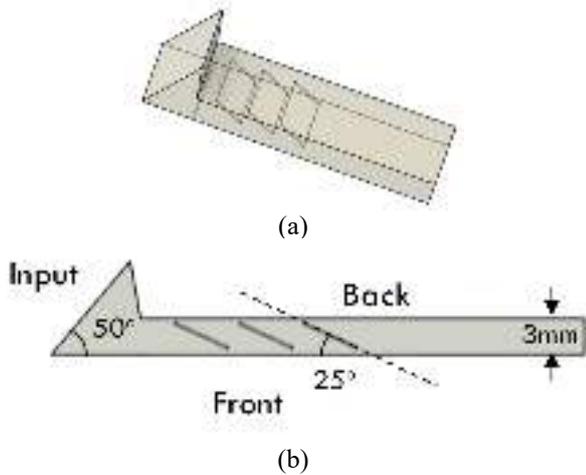

(a)

(b)

Fig. 2 (a) A picture illustrating the overall architecture of the proposed near-field AR display with 3 reflectors, and (b) the corresponding side view of the AR display fabrication process. By 3D-printing a custom-designed mold, we can fabricate our AR display to a desired shape. The benefits of using 3D-printed-mold include: i) ease of manufacturing, ii) cost-effective, and iii) fast fabrication. Based on the simulation result, we 3D-printed our modular mold using polymer-resin. During the fabrication, three glass-slips have been embedded as light-reflectors. Each glass-slip is dielectric coated with a reflectivity of 10% across the visible wavelength. In our prototype, we used the PDMS Sylgard™ 184 produced by Dow Corning. It consists of two parts – a silicon base and a curing agent. Depending on the ratio between the base and the curing agent, the toughness can be controlled [11]. The well-mixed PDMS is gently poured into the modular-mold. Afterwards, the mold is degassed in a vacuum chamber and cured.

The AR display is extracted from the mold after the curing. All glass-slips have been positioned at the expected locations. The insert of Fig. 3 shows the final prototype of our proposed flexible near-field AR display.

### IV. Optical Characterisation of the AR display

In our experimental arrangement, as illustrated in Fig. 3, we have devised a bespoke configuration that enables us to precisely orient a commercially accessible light engine (microdisplay) at a 90-degree angle relative to the waveguide's coupling edge. To maintain the waveguide's stability, we have employed 3D-printed structures to secure it on a transition stage. This arrangement not only guarantees the effective transmission of light through the waveguide, an essential element in the immersive AR environment we are actively developing but also grants us a full view of the waveguide's entire active region.

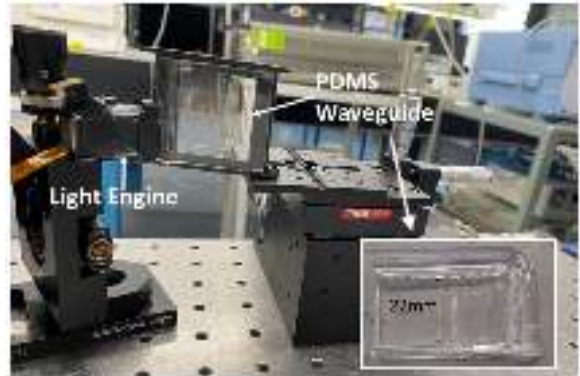

Fig 3: Illustration of the experimental setup, demonstrating alignment of the PDMS AR-display with a light engine coupled at the edge. An off-the-shelf light engine is affixed perpendicularly to the waveguide's coupling face. The insert is our PDMS AR display with 3 embedded reflectors.

### V. Results and discussion

We launch different images and videos into the prototype and we capture the image exiting from the AR display. Figure 4 (a) and (b) show the images that have been launched into the AR display and the captured image from the output of the AR display, respectively.

From the experimental results, we can confirm that the PDMS based near-field AR display is working as expected. The input image has been reconstructed by the AR display. The image is clear enough for human to identify majority of the features in the input image. It is the first demonstration of a flexible AR display using PDMS as the optical combiner and waveguide.

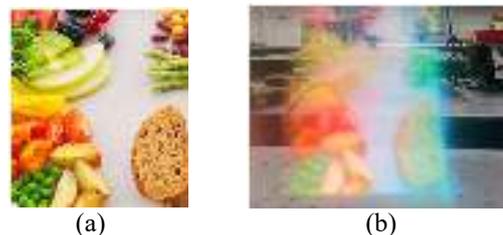

(a)          (b)

Fig. 4 (a) the image launched into the near-field AR display, and (b) the corresponding output image superimposed with real world through the AR display.

### VI. Summary

We have prototyped a near-eye flexible AR display using PDMS. Different images and videos have been launched into the lightguide and the images have been reconstructed with good quality. The successful experimental demonstration of the PDMS based near-field AR display showed that our proposed design and fabrication method are able to realize the required operation constraints of a practical near-field AR display. To the best of our knowledge, it is the first demonstration of a flexible near-eye AR display, which is superior to glass-based AR display for their ability to withstand physical stress. Our proposed manufacturing technique is cost-effective and holds the potential for mass production.

ACKNOWLEDGMENT

This work was conducted in part at the Melbourne Centre for Nanofabrication (MCN) in the Victorian Node of the Australian National Fabrication Facility (ANFF). Additionally, this project was supported by Research Funding from JARVISH.

REFERENCES

[1] Carmigniani, J. and Furht, B., "Augmented reality: an overview," Handbook of augmented reality, pp.3-46, 2011.
[2] Nee, A.Y., S.K. Ong, G. Chryssolouris, and D. Mourtzis, "Augmented reality applications in design and manufacturing," CIRP annals, 61(2), pp.657-679, 2012.
[3] T. Langlotz, T. Nguyen, D. Schmalstieg and R. Grasset, "Next-generation augmented reality browsers: rich, seamless, and adaptive," Proceedings of the IEEE, 102(2), pp.155-169, 2014.
[4] J. Xiong, E. L. Hsiang, Z. He, T. Zhan, and S. T. Wu, "Augmented reality and virtual reality displays: emerging technologies and future perspectives. Light: Science & Applications, 10(1), p.216, 2021.
[5] T. Zhan, J. Xiong, J. Zou, and S. T. Wu, "Multifocal displays: review and prospect," PhotoniX, 1, pp.1-31, 2020.
[6] B. C. Kress, I. Chatterjee ,"Waveguide combiners for mixed reality headsets: a nanophotonics design perspective," Nanophotonics, 10(1), pp.41-74, 2020.
[7] Y. H. Lee, T. Zhan, and S. T. Wu, "Prospects and challenges in augmented reality displays," Virtual Real. Intell. Hardw., 1(1), pp.10-20, 2019.
[8] R Rajasekharan, Q Dai, TD Wilkinson, Electro-optic characteristics of a transparent nanophotonic device based on carbon nanotubes and liquid crystals, Applied optics 49 (11), 2099-2104, 2010
[9] R. Rajasekharan, T. D. Wilkinson, P. J. W. Hands, and Q. Dai, "Nanophotonic Three-Dimensional Microscope," Nano Lett. 2011, 11, 7, 2770–2773, 2011.
[10] T. Zhan, K. Yin, J. Xiong, Z. He, and S. T. Wu, "Augmented reality and virtual reality displays: perspectives and challenges," Iscience, 23(8), 2020.
[11] Zhixin Wang, A. V. Alex, D. G. Nathan, "Crosslinking effect on polydimethylsiloxane elastic modulus measured by custom-built compression instrument," J. Appl. Polym. Sci., 131,41050, 2014.
[12] M. Sun, W. Shieh, and R. Unnithan, "Design of Plasmonic Modulators with Vanadium Dioxide on Silicon-On-Insulator," IEEE Photonics Journal, vol. 9, no. 3, pp. 1-10 ,2017.